\documentclass{emulateapj}
\usepackage{mycommands}
\usepackage{verbatim}
\usepackage{graphicx}
\usepackage[percent]{overpic}
\usepackage{subfigure}
\usepackage{esint}
\usepackage{url}
\begin{document}


\title{{\it PICsar}: A 2.5D Axisymmetric, Relativistic, Electromagnetic, Particle in Cell Code with a Radiation Absorbing Boundary}

\begin{abstract}
We present {\it PICsar} -- a new Particle in Cell code geared towards efficiently simulating the magnetosphere of the aligned rotator. {\it PICsar} is a special relativistic, electromagnetic, charge conservative code that can be used to simulate arbitrary electromagnetics problems in axisymmetry. It features stretchable body-fitted coordinates that follow the surface of a sphere, simplifying the application of boundary conditions in the case of the aligned rotator; a radiation absorbing outer boundary, which allows a steady state to be set up dynamically and maintained indefinitely from transient initial conditions; and algorithms for injection of charged particles into the simulation domain. The code is parallelized using MPI and scales well to a large number of processors. We discuss the numerical methods used in {\it PICsar} and present tests of the code. In particular, we show that {\it PICsar} can accurately and efficiently simulate the magnetosphere of the aligned monopole rotator in the force free limit. We present simulations of the aligned dipole rotator in a forthcoming paper.
\end{abstract}

\author{Mikhail A. Belyaev}
\affil{Astronomy Department, University of California,
    Berkeley, CA 94720}

\section{Introduction}
\label{intro}

The modeling of pulsar magnetospheres has a  history dating back to the association of pulsars with spinning neutron stars. One of the early magnetospheric models that remains relevant today is that of \citet{GoldreichJulian}, who studied a magnetosphere filled with plasma in the special case when the neutron star rotational and magnetic axes are aligned (aligned dipole rotator). The \citet{GoldreichJulian} model assumes a``force-free" magnetosphere, which means that electromagnetic fields dominate particle inertia so that the Lorentz force law for the plasma is $\rho \bfE + c^{-1}\bfJ \times \bfB = 0$, where $\rho$ is the charge density, $\bfJ$ is the current, and $c$ is the speed of light. The force-free condition implies that $\bfE \cdot \bfB = 0$, so there is enough plasma to short out the component of the electric field along magnetic field lines. The force-free approximation can be viewed as a limiting case of relativistic magnetohydrodynamics (MHD) when particle pressure and inertia are negligible \citep{KomissarovFFE}.

The structure of the pulsar magnetosphere has been simulated using force-free and MHD simulations for both the aligned rotator \citep{Contopoulos,Gruzinov,Timokhin_forcefree,McKinney,Komissarov}, and for arbitrary inclinations of the spin and magnetic axes \citep{Spitkovsky,CK,Kalapotharakos_ext,Tchekhovskoy} . These simulations were extended to include a finite resistivity by \citet{LiSpitkovsky,Kalapotharakos}, which allowed the authors to capture a spectrum of solutions between the force-free and vacuum limits. One of the observational impetuses for more accurate solutions of magnetospheric structure is the modeling of pulsar lightcurves \citep{BaiSpitkovsky,Kalapotharakos_lightcurve}.

One limitation of force-free, MHD, and even the resistive simulations is that they don't address the creation and acceleration of particles in the magnetosphere. These particles are created via pair-production and accelerated in vacuum gaps \citep{Sturrock,AronsScharlemann,ChengHoRuderman,HardingMuslimov}, where the force-free assumption breaks down ($\bfE \cdot \bfB \ne 0$). To model the effect of pair production on the global structure of the magnetosphere and gain a deeper understanding of magnetospheric emission, it is natural to perform particle-based simulations. 

A well-developed technique for electromagnetic simulation using particles is the Particle in Cell (PIC) method e.g. \citet{BirdsallLangdon}. The major reason for using PIC is that it offers a kinetic and self-consistent approach to the solution of Maxwell's equations in a plasma. The PIC method has already been used to simulate physics relevant to pulsar magnetospheres including pair production and particle acceleration in vacuum gaps \citep{Timokhin_paircascades,TimokhinArons} and instabilities in electron-positron current sheets \citep{ZenitaniHoshino2007,ZenitaniHoshino2008}. These simulations were local, however, and it is our aim to model the pulsar magnetosphere using global PIC simulations.

Recently, \citet{PhilippovSpitkovsky} have simulated the magnetosphere of the aligned rotator using a relativistic, electromagnetic 3D Cartesian PIC code and found that their results are consistent with force-free simulations to $\sim 10\%$. However, Cartesian simulations of pulsar magnetospheres are prohibitively expensive to perform in terms of computational cost. The fundamental reason for this is that Cartesian coordinates are far from ideal for simulating a physical system, which is most naturally described in spherical geometry. \citet{WadaShibata} proposed a purely particle-based approach (not PIC) that did away with the computational mesh altogether. However, their method is electrostatic at its core, which limits its applicability. Even more recently, \citet{chenbeloborodov} performed 2.5D axisymmetric PIC simulations of the pulsar magnetosphere. They found that pair production all the way out to the light cylinder was necessary to generate a force-free magnetosphere, showing that global PIC modeling of the pulsar magnetosphere can lead to fundamental insights and help test theoretical models.

Our aim in this paper is to demonstrate the efficiency and accuracy of PIC simulations for modeling the pulsar magnetosphere. Although we present actual simulations of the aligned dipole rotator in a forthcoming paper, the new {\it PICsar} code presented here is ideally suited to simulating the magnetosphere of the aligned rotator. {\it PICsar} is a 2.5D axisymmetric, relativistic, electromagnetic, charge conservative PIC code that is several orders of magnitude faster for axisymmetric pulsar simulations than a 3D Cartesian PIC code. 

{\it PICsar} implements coordinates that are body-fitted to the surface of a sphere. This makes the boundary condition on the surface of the star simple to implement and eliminates the need for simulating the plasma in the neutron star interior. Furthermore, {\it PICsar} implements a radiation absorbing outer boundary condition, which allows electromagnetic waves to leave the simulation domain and set up a steady state dynamically.

This paper is organized as follows. In \S \ref{overview} we present an overview of {\it PICsar}, and in \S \ref{algsec} we give a detailed account of the algorithms used in {\it PICsar}. In \S \ref{testsec} we present tests of {\it PICsar}, which demonstrate the accuracy of the code and its capabilities. In \S \ref{monosec} we present simulations of the aligned monopole rotator and present an algorithm for charge injection. Finally, we discuss our results in \S \ref{discussion_sec}. 

\section{{\it PICsar} Overview}
\label{overview}

{\it PICsar} is a 2.5D axisymmetric, relativistic, electromagnetic, charge conservative PIC code with a radiation absorbing outer boundary. It is based on the 3D Cartesian PIC code TRISTAN \citep{TRISTAN}, but has been heavily modified to work in 2.5D axisymmetry and has been parallelized using MPI. 

A PIC code is a type of particle-mesh code in which electric and magnetic fields are stored and updated on a grid using Maxwell's equations \citep{BirdsallLangdon}. The moniker ``2.5D axisymmetric" means that there are in general six nonzero field components (three of electric and three of magnetic field) at each point, but that the azimuthal derivative, $\partial/\partial \phi$, of any field quantity is equal to zero. From a computational point of view, this means that a 2D grid ($r-\theta$) of 3D vectors needs to be simulated, which explains the use of the term 2.5D. 

In an electromagnetic PIC code, the equations solved to advance the fields from one time step to the next are the time-dependent Maxwell's equations. These are written most compactly in Lorentz-Heaviside units, and we will use Lorentz-Heaviside units throughout, unless explicitly stated. The time-dependent Maxwell's equations in Lorentz-Heaviside units are
\ba
\label{maxtd}
\frac{\partial \bfB}{\partial t} &=& -c\bfnabla \times \bfE  \\
\frac{\partial \bfE}{\partial t} &=& c\bfnabla \times \bfB - \bfJ \nn,
\ea
and the time-independent Maxwell equations are
\ba
\label{maxti}
\bfnabla \cdot \bfE &=& \rho \\
\bfnabla \cdot \bfB &=& 0 \nn.
\ea

Only the time-dependent Maxwell equations are solved by the code, and the time-independent equations form a pair of constraints. The Yee algorithm \citep{Yee}, which is second order in both space and time is used to update the E and B fields and ensures that if the constraint equations (\ref{maxti}) are satisfied initially, then they are satisfied for all time (to machine precision) in the absence of sources (i.e. charged particles). 

In order to satisfy equations (\ref{maxti}) in the presence of sources, the current deposited to the numerical grid must satisfy the equation of charge conservation. If the only sources are particles and the $n$-th particle has a shape function (i.e. spatial charge distribution) given by $\rho_n(\bfx-\bfx_n)$, where $\bfx_n$ is the particle position, then the equation of charge conservation is
\ba
\label{Jcons}
\sum_n\frac{\partial \rho_n}{\partial t} = -\bfnabla \cdot \bfJ.
\ea
For charge conservative current deposition in {\it PICsar}, we use the Villasenor-Buneman algorithm \citep{VillasenorBuneman}, which takes the particles to have the same shape function as the grid cells. Thus, in a curvilinear coordinate system the shape function of the particles is not constant, but changes to reflect the local shape and stretch of the coordinate system.

From equation (\ref{Jcons}), it is clear that as the particles move around the grid, they deposit current. The job of the particle mover in a PIC code is to move and accelerate the particles from one timestep to the next. In a relativistic, electromagnetic PIC code, particle motion is dictated by the Lorentz force law
\ba
 \frac{d \gamma m \bfv}{d t} = q(\bfE + \frac{\bfv}{c} \times \bfB).
\ea 
{\it PICsar} implements a choice of two different algorithms for the mover step in the code: the Boris \citep{Boris} and Vay \citep{Vay} algorithms. 

One important advantage of the Vay algorithm over the Boris algorithm as concerns pulsar simulation is that the Vay algorithm correctly captures the average $E \times B$ particle drift motion regardless of the value of $\omega_c \Delta t$, where $\omega_c$ is the cyclotron frequency and $\Delta t$ is the simulation timestep. Due to the small Larmor radius of the particles in the pulsar magnetosphere, the $E \times B$ drift is dominant over other particle drifts and capturing it correctly is essential to the accuracy of the simulation. 

The final major feature of {\it PICsar} is the radiation absorbing outer boundary, which is a finite width lossy layer that exponentially attenuates electromagnetic waves passing through it. This allows a steady state solution to be set up and maintained indefinitely starting from transient initial conditions.

\section{Numerical Methods}
\label{algsec}
Having outlined {\it PICsar}, we now give a more detailed description of the algorithms it implements. 

\subsection{Field Solve}
\label{fieldsolvesec}
We begin by describing the update step for the fields in 2.5D axisymmetry. With the exception of geometrical factors introduced by the curvilinear coordinates, the field solve in {\it PICsar} is identical to the one in the TRISTAN and follows the Yee algorithm \citep{Yee}. 
 
In the Yee algorithm, the $E$ and $B$ 3D vector fields are offset from each other in time by half a timestep. This allows for second order convergence in time by ``leapfrogging" the $E$ and $B$ fields. The components of the $E$ and $B$ vector fields are offset from each other by half a grid cell in each spatial direction. The current, $J$, which is also a 3D vector field, is deposited by the particles to the grid and has vector components defined at the same grid positions as the vector components of the $E$ field. In {\it PICsar}, the 3 vector components of the $E$, $B$, and $J$ fields are located on the 2D grid at positions
\begin{align}
\label{offseteq}
E_r^{ij}, \ J_r^{ij} &\rightarrow  (i+.5, \ j)  &B_r^{ij} &\rightarrow  (i, \ j+.5) \\
E_\theta^{ij}, \ J_\theta^{ij} &\rightarrow (i, \ j+.5) &B_\theta^{ij} &\rightarrow (i+.5, \ j) \nn \\
E_\phi^{ij}, \ J_\phi^{ij} &\rightarrow (i, \ j) &B_\phi^{ij} &\rightarrow (i+.5, \ j+.5). \nn
\end{align}
Here $i,j$ are integer grid coordinates, and the pair of values to the right of the arrow indicates the grid location of the vector component. 

\begin{figure}[!t]
\centering
\includegraphics[width=.49\textwidth]{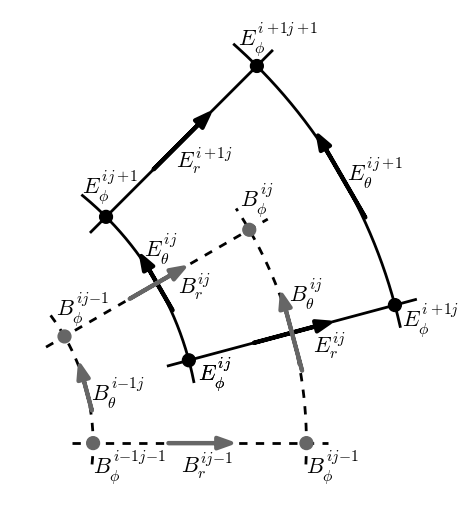}
\caption{Relative locations of electric and magnetic field vectors on the computational grid.}
\label{vec_grid_fig}
\end{figure}

Fig \ref{vec_grid_fig} is a diagram showing the locations of the $E$ and $B$ fields on the grid. The solid curves outline a grid cell bounded by the grid points $(i,j)$, $(i+1,j)$, $(i+1,j+1)$, $(i,j+1)$. The black vectors and circles show the locations of $E$ field components on the grid. The dashed curves outline a``dual cell" which is bounded by the half-integer grid points $(i-.5,j-.5)$, $(i+.5,j-.5)$, $(i+.5,j+.5)$, $(i-.5,j+.5)$. The grey vectors and circles show the locations of the $B$ field components on the grid. 

Offsetting the electric and magnetic fields in space according to equation (\ref{offseteq}) allows Maxwell's equations to be solved in conservative form on the discretized grid. The mathematical details of the procedure for advancing the fields in time are given in Appendix \ref{field_solve_app}. 

\subsection{Coordinate Systems}
\label{coordsec}
{\it PICsar} is capable of solving Maxwell's equations in axisymmetry in any coordinate system which has grid lines that are parallel to the constant $\theta$ and constant $r$ grid lines of the standard spherical coordinate system. Such a coordinate system is ``body-fitted" to the surface of a sphere but allows for stretching in the radial and meridional directions.

To make this notion more concrete, we define the radial and meridional stretching functions $f_r(i)$ and $f_\theta(j)$, where $(i,j)$ is a grid coordinate. These functions give the $(r,\theta)$ coordinate of the $(i,j)$ grid-point. Choosing linear functions for $f_r$ and $f_\theta$ yields the spherical coordinate system. Any non-linear choice for $f_r$ and $f_\theta$ results in a coordinate system that is stretched with respect to spherical coordinates, but which has grid lines that are parallel to the lines of constant $r$ and constant $\theta$. 

\begin{figure}[!t]
\centering
\includegraphics[width=.49\textwidth]{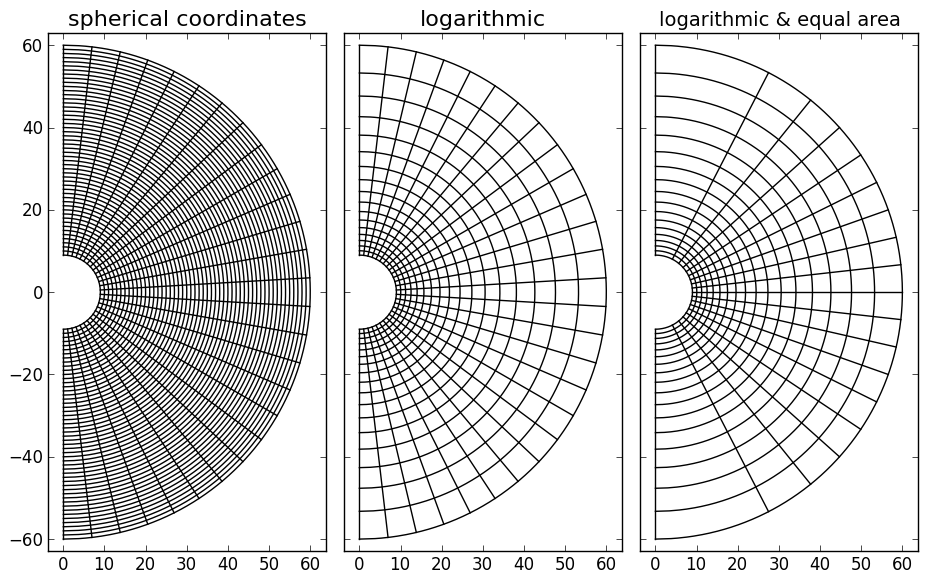}
\caption{Comparison of equally spaced grid lines in three different coordinate systems. From left to right: spherical coordinates $(r,\theta)$,  logarithmic radial coordinates $(\xi,\theta)$, logarithmic radial coordinates and equal area meridional coordinates $(\xi,\theta_A)$.}
\label{coordfig}
\end{figure}

{\it PICsar} implements two different options for both $f_r$ and $f_\theta$. The options for $f_r$ are linear and logarithmic and the options for $f_\theta$ are linear and ``equal area." As already mentioned, choosing linear stretching functions in both the radial and meridional directions results in the standard spherical coordinate system. We now discuss logarithmic and equal area coordinates, and their advantages over spherical coordinates for PIC simulations. Fig \ref{coordfig} shows the grid lines for these coordinate systems in comparison to spherical coordinates.

\subsubsection{Logarithmic Coordinates in the Radial Direction}
The logarithmic coordinate in the radial direction, $\xi$, is given by the transformation
\ba
\xi \equiv r_0 \ln \left(r/r_0\right).
\ea
This has the advantage that the resolution is enhanced close to the surface of the neutron star, and a much larger grid can be simulated in the radial direction compared to spherical coordinates. Moreover, a logarithmic coordinate system is special in that it preserves the aspect ratio (i.e. $\Delta r/r \Delta \theta$) of the grid cells as a function of radius. Thus, the cells do not become elongated in the meridional direction with increasing radius as occurs in spherical coordinates.

\subsubsection{Equal Area Coordinates in the Meridional Direction} 
The equal area coordinate in the meridional direction, $\theta_A$, is given by the transformation
\ba
\label{thetaA}
\theta_A \equiv -\cos \theta
\ea 
and is defined on the range $[-1,1]$. At constant radius, any two points separated by the same value of $\Delta \theta_A$ enclose the same surface area, $\Delta A = 2 \pi \Delta \theta_A$.

As we show in \S \ref{noisypolar}, using equal area coordinates in the meridional direction greatly reduces the particle noise in the simulation. This is because the charge density of an individual particle is greatly enhanced at the poles versus the equator in spherical coordinates for a particle shape function that conforms to the shape of the numerical grid. On the other hand, the charge density is constant in the meridional direction in equal area coordinates, which greatly reduces particle granularity near the poles. 

\subsection{Particle Mover and Current Deposit}
\label{movdepsec}

The particle mover applies an acceleration to the particle velocities and moves the particles within the simulation domain. Although {\it PICsar} uses a curvilinear coordinate system, the actual particle acceleration and push steps are carried out in Cartesian coordinates using the standard Vay \citep{Vay} or Boris \citep{Boris} algorithms. This simply requires transforming between curvilinear and Cartesian coordinates in the mover.

The current deposit step requires no modification in going from Cartesian to curvilinear coordinates when using the charge conservative Villasenor-Buneman algorithm \citep{VillasenorBuneman}. This is in fact true not only of the Villasenor-Buneman algorithm, but of any current deposit algorithm that only references grid coordinates (e.g. $(i,j)$) and makes no reference to physical coordinates (e.g. $(x,y)$, $(r,\theta)$, etc.). For any such algorithm, the particle shape function either conforms to the shape of the grid or is a delta function. Thus, by keeping the Villasenor-Buneman algorithm without modification, we are simply stating that the particles have a shape function which is a spherical annulus in axisymmetric coordinates rather than a rectangular prism as in Cartesian coordinates. 

In order to reduce noise in the simulation, we filter the current deposited to the grid by the particles. The reason for this is that fluctuations in the current due to small number particle statistics are a major source of noise in PIC simulations. Particle noise only decreases as $\propto 1/\sqrt{N}$, where $N$ is the number of particles in the simulation, and thus it is typically too computationally expensive to eliminate it by increasing the number of particles. Current filtering damps the high frequency harmonics in the simulation in an inexpensive way by spatially spreading the particle charge shape function. It can be applied any number of times, $N_\text{filt}$, to the current, and the code remains charge conservative. The only modification is that with current filtering, the particle shape function takes the form of the original unfiltered shape function with the filter applied to it $N_\text{filt}$ times. This is because filtering the current after it is deposited to the grid is equivalent to filtering the particle shape function and then depositing the current to the grid. Appendix \ref{filterapp} gives a more in-depth treatment of current filtering in {\it PICsar}.

After the current has been deposited to the grid and filtered (if desired), it can be used to update the components of the electric field on the grid (see Appendix \ref{field_solve_app} for details.)

\subsection{Boundary Conditions}

\subsubsection{Field Boundary Conditions}
\label{fbc}
At the inner and outer radial boundaries, we implement conducting boundary conditions. For a perfect conductor with zero field inside, $\bfE_\parallel = 0$ and $\bfB_\perp = 0$, so we simply set $E_\theta = E_\phi = B_r = 0$ on the radial boundaries. Note that according to equations (\ref{offseteq}), $E_\theta$, $E_\phi$, and $B_r$ are all located at integer radial grid coordinates, whereas $B_\theta$, $B_\phi$, and $E_r$ are located at half integer radial grid coordinates. Thus, $B_\theta$, $B_\phi$ and $E_r$ do not lie on the radial boundaries, and thus no boundary condition in the radial direction needs to be applied for these field components. Boundary conditions and update equations for the fields on the polar axis are discussed in Appendix \ref{field_solve_app}.

The inner and outer conducting boundaries trap electromagnetic waves within the simulation domain. In order to absorb radiation, {\it PICsar} implements a finite width lossy layer just inside the outer boundary. This lossy layer has both a finite electrical conductivity, $\sigma$, and a finite magnetic conductivity, $\sigma_m = \sigma$. The magnetic conductivity is artificial, but induces exponential damping of electromagnetic waves as they propagate through the lossy layer (Chapter 3.12 in \citet{fdtdbook}). Equations (\ref{maxtd}) with the magnetic and electric conductivities included are given by
\ba
\frac{\partial \bfB}{\partial t} &=& -c\bfnabla \times \bfE - \sigma \bfB  \\
\frac{\partial \bfE}{\partial t} &=& c\bfnabla \times \bfB - \bfJ -\sigma \bfE \nn.
\ea

The magnetic and electric conductivities can be implemented numerically by simply setting
\ba
\label{EBcond}
\bfE^{ij} = (1-\sigma^*)\bfE^{ij} \\
\bfB^{ij} = (1-\sigma^*)\bfB^{ij} \nn
\ea
inside the lossy layer after the regular field update and deposit steps have been performed by the code. Note, that we have used $\sigma^*$ in equations (\ref{EBcond}) to parametrize the numerical value of $\sigma$ and that by design $0 \le \sigma^* \le 1$. 

For significant attenuation, we find that the width of the lossy layer, $\Delta L$, should be at least 10 cells; keep in mind that the wave is doubly attenuated, first as it passes through the lossy layer in the outward direction and again after it reflects off the outer conductor and passes through the lossy layer in the inward direction. As for $\sigma^*$, we find that a value of $\sigma^*$ that increases linearly with radius from 0 at the inner edge of the lossy layer to $2/\Delta L$ at the outer edge results in good attenuation. The reason for the linear profile of $\sigma_*$ is to better match the impedance between the simulation domain and the lossy layer and to minimize reflections off the inner edge of the lossy layer.

\subsubsection{Particle Boundary Conditions}
\label{pbc}

We now discuss boundary conditions on the particles in the radial and meridional directions. The meridional boundaries of the simulation domain lie on the polar axis. In axisymmetry, the correct boundary condition on the polar axis is a reflection boundary condition. Thus, particles cannot leave the simulation domain in the meridional direction.

Particles can and do leave the simulation domain through the inner and outer radial boundaries. A particle which passes outside the simulation domain in the radial direction and hits either the inner or outer conductor should be deleted and removed from the simulation. However, a particle cannot be deleted immediately once it hits a conductor, since a fraction of its charge will still be ``hanging" outside the conductor and inside the simulation domain due to the finite width of a particle shape function. 

A simple solution to this problem that we use in our pulsar simulations is to only delete a particle once its entire shape function lies inside the conductor. In this way, all of the particle charge is hidden behind the conductor and no charge is left hanging within the simulation domain when the particle is deleted. This method works even with current filtering, although a particle must be at least $N_\text{filt}+1$ cells inside the edge of the conductor before its shape function lies entirely within the conductor and it can be deleted. Particles within the inner conductor move according to the corotation $E \times B$ field, while those within the outer conductor fly ballistically, moving on straight lines with no acceleration. 

\section{Tests of {\it PICsar}}
\label{testsec}

\subsection{Second Order Convergence}
\label{2ndordersec}
We begin by demonstrating the second order convergence of the field solve step, which was described in \S \ref{fieldsolvesec}. For this test, we turn off the radiation absorbing outer layer (\S \ref{fbc}), so we have conducting boundary conditions for both the inner and outer radial boundaries. In the space between the conductors, we set up an axisymmetric spherical transverse magnetic (TM) mode (\citet{Jackson} Section 8.9), which has the solution 
\ba
\label{TMmodeeq}
B_\phi(r,\theta,t) &=& B_0 \frac{u_l(k r)}{k r}P_l^1(\cos \theta)e^{-i \omega t} \\
E_r(r,\theta,t) &=& \frac{i}{k r \sin \theta}\frac{\partial}{\partial \theta}(\sin \theta B_\phi) \nn \\
E_\theta(r,\theta,t) &=& -\frac{i}{k r}\frac{\partial}{\partial r}(r B_\phi) \nn,
\ea 
where $u_l(r)/r$ are spherical Bessel functions, $P_l^1$ are the associated Legendre polynomials of first order, and $\omega = c k$. The conducting boundary condition $E_\theta = 0$ constrains the value of $k$ such that $du_l/dr = 0$ on the inner and outer boundaries of the simulation domain. 

For our test, we take $l=1$, in which case the special functions become
\ba
u_1(r) = \frac{\sin(r)}{r} - \cos(r), \ \ P_1^1(\cos \theta) = -\sin \theta.
\ea
We take the radius of the inner conductor to be the first root and the radius of the outer conductor to be the fourth root of $du_1/dr$, which ensures $E_\theta = 0$ on the conducting boundaries. We take the start of the simulation to be at $t=0$, in which case $\text{Re}[E_r(r,\theta,0)] = \text{Re}[E_\theta(r,\theta,0)] = 0$, where Re denotes the real part. 

\begin{figure}[!t]
\centering
\includegraphics[width=.49\textwidth]{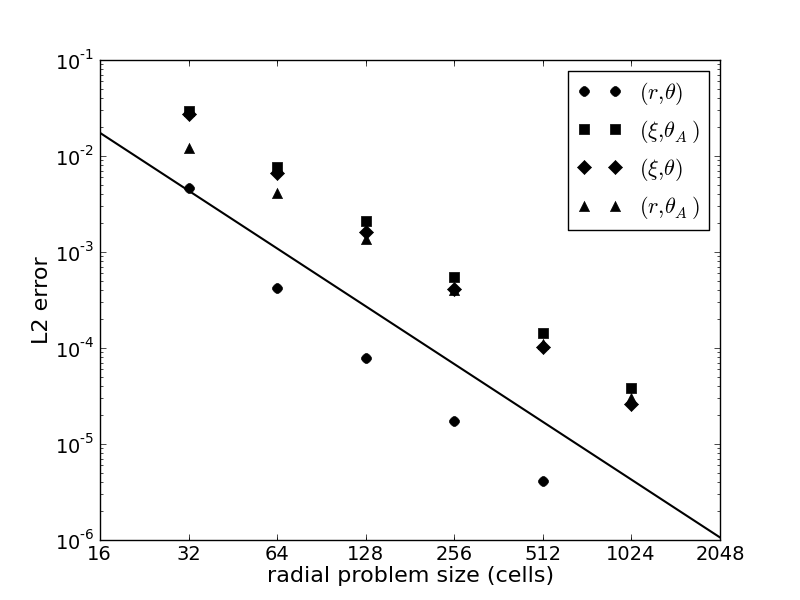}
\caption{Log-log plot of the L2 error between the numerical and analytical solutions at time $t=1.25/\omega$ as a function of resolution.}
\label{TMmodefig}
\end{figure}

Because $B$ lags $E$ by half a timestep, we must initialize the $B_\phi$ field at time $t = -\Delta t/2$. The $B_r$, $B_\theta$, and $E_\phi$ fields are initially zero and remain so for all time. We allow the simulation to run for a time $t = 1.25/\omega$ and then compute the error in the L2 norm for $rB_\phi$, which is defined as
\ba
\mathcal{L}_2 \equiv \frac{\sqrt{\sum\limits_{i=1}^{N_r} \sum\limits_{j=1}^{N_\theta} f_r(i+1/2)^2 \left(B_\phi^{i j}-\widetilde{B}_\phi^{i j}\right)^2}}{\sqrt{\sum\limits_{i=1}^{N_r} \sum\limits_{j=1}^{N_\theta} f_r(i+1/2)^2 \left(\widetilde{B}_\phi^{i j}\right)^2}},
\ea
where $B_\phi$ is the numerical solution, $\widetilde{B}_\phi$ is the analytic solution, and the radial stretching function, $f_r$, was introduced in \S \ref{coordsec}. We plot the L2 error as a function of resolution on a log-log plot in Fig \ref{TMmodefig} for four different coordinate systems -- $(r,\theta)$ (circles), $(\xi,\theta_A)$ (squares), $(\xi, \theta)$ (diamonds), and $(r,\theta_A)$ (triangles). The line has the proper slope for second order convergence, and all four coordinate systems converge at second order. Although the absolute error is smallest for spherical coordinates in this particular test, spherical coordinates are actually the most susceptible to ringing when particles are present near the polar axis.

\subsection{Charged Sphere Test}
\begin{figure*}[!t]
\centering
\subfigure{\begin{overpic}
		  [height=.25\textheight,trim=0cm 0cm 2.2cm 0cm,clip=true]{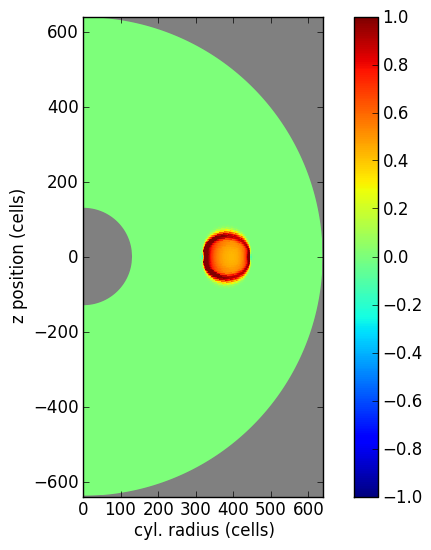}
		  \put(17,91){\normalsize a}
		  \put(41,89){\normalsize$.125t_c$}
		  \end{overpic}}
\subfigure{\begin{overpic}
		  [height=.25\textheight,trim=2.0cm 0cm 2.2cm 0cm,clip=true]{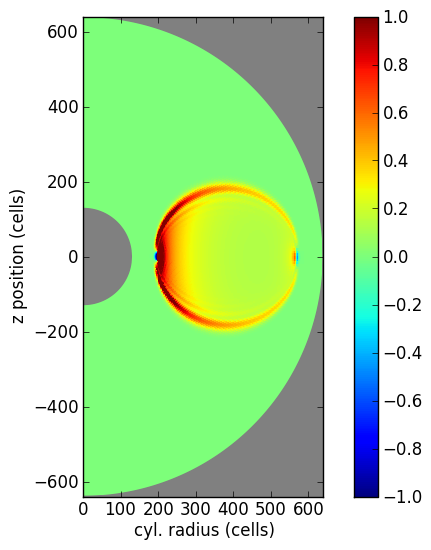}
		  \put(3,91){\normalsize b}
		  \put(28,89){\normalsize$.375 t_c$}
		  \end{overpic}}
\subfigure{\begin{overpic}
		  [height=.25\textheight,trim=2.0cm 0cm 2.3cm 0cm,clip=true]{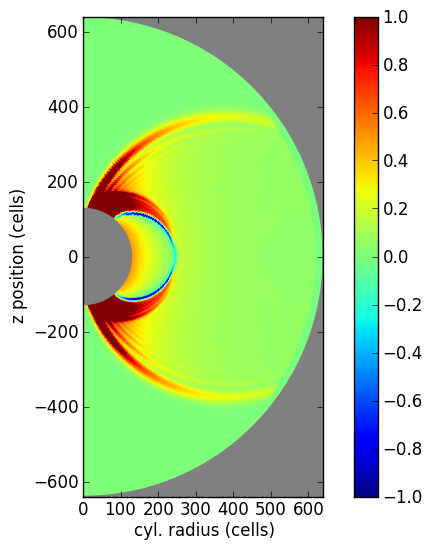}
		  \put(3,91){\normalsize c}
		  \put(29,89){\normalsize$.75 t_c$}
		  \end{overpic}}
\subfigure{\begin{overpic}
		  [height=.25\textheight,trim=2.0cm 0cm 2.3cm 0cm,clip=true]{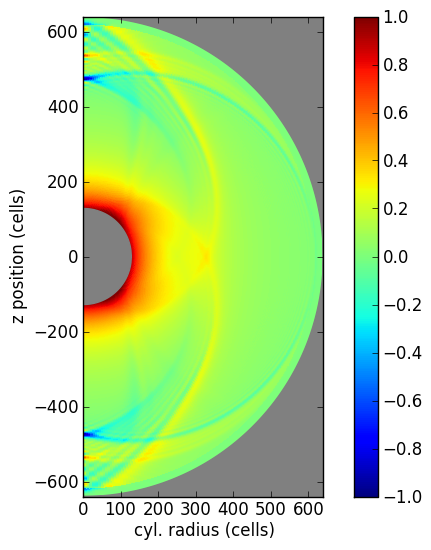}
		  \put(3,91){\normalsize d}
		  \put(29,89){\normalsize$1.5 t_c$}
		  \end{overpic}}
\subfigure{\begin{overpic}
		  [height=.25\textheight,trim=2.0cm 0cm 0cm 0cm,clip=true]{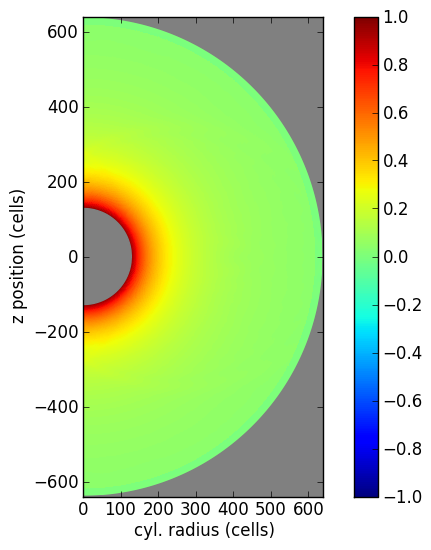}
		  \put(3,91){\normalsize e}
		  \put(29,89){\normalsize$3.5 t_c$}
		  \end{overpic}}
\caption{Sequence of images showing $E_r$ that depicts how the inverse square radial electric field is established dynamically for the simulation described in \S\ref{statdynfield}. A positive charge is shot radially inward into the central conductor along the equator and its negative pair is shot radially outward toward the outer conductor also along the equator. The pair is initialized halfway between inner and outer conductor. The time in units of the light crossing time, $t_c$, is shown in the upper right corner.}
\label{chargedynfig}
\end{figure*}

\begin{figure*}[!t]
\centering
\subfigure{\includegraphics[width=.45\textwidth]{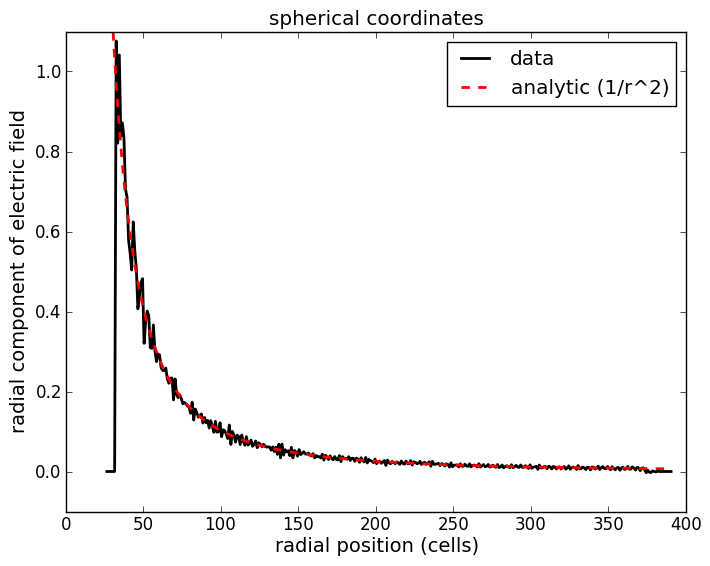}}
\subfigure{\includegraphics[width=.45\textwidth]{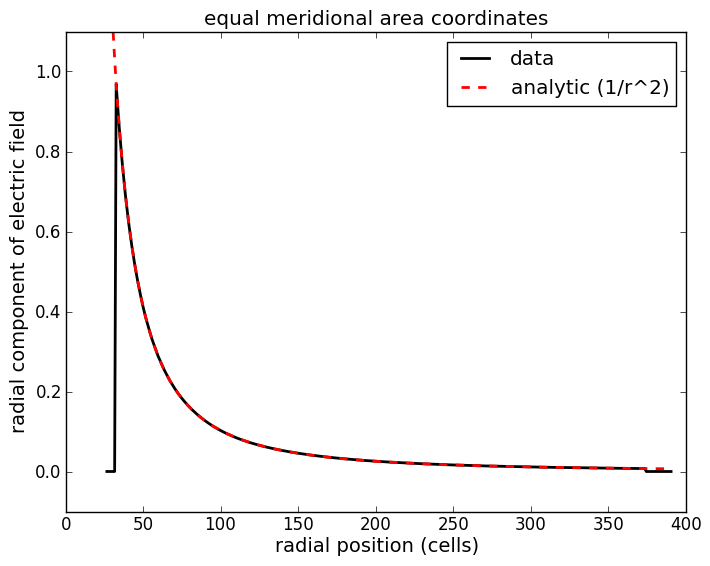}}
\caption{Plot of the radial electric field $E_r$ after $t=3.5t_c$ along the polar axis for simulation described in \S\ref{noisypolar}. Dashed red line -- analytic solution (inverse square law). Black solid line -- simulation. Left panel: simulated field in spherical coordinates shows excessive ringing. Right panel: simulated field is smooth in equal meridional area coordinates.}
\label{noisyfig}
\end{figure*}

Our next test is establishing an inverse square radial electric field dynamically. We perform this test with a charge falling into the inner conductor along both the equator and the polar axis.

\subsubsection{Charge Falls in Along Equator}
\label{statdynfield}
We show how a static field is set up dynamically and demonstrate the effectiveness of the radiation boundary conditions at damping electromagnetic waves by having a charge fall into the inner conductor in the equatorial plane, while its partner charge moves outward and hits the outer conductor. Note that due to the axisymmetry of the coordinate system, the charge is not a point charge as in 3D, but rather an annulus. We use the radiation absorbing outer boundary (\S \ref{fbc}), which is essential for absorbing the radiation from the transient. We also filter the current (\S \ref{movdepsec}) in order to damp the high frequency harmonics, which induce ringing and are not well-resolved on the grid. 

Our setup consists of a positive and a negative charge both initially located in the equatorial plane halfway between the inner and outer conductors. The positive charge is shot radially inward at a velocity $v_r = -.95c$, and the negative charge is shot radially outward at a velocity of  $v_r = .95c$. 

After about $.5 t_c$, where $t_c$ is the light crossing time across the simulation domain, the positive/negative charge has fallen into the inner/outer conductor and has been deleted from the simulation. After the positive charge falls into the inner conductor, an electromagnetic pulse is launched that propagates along the surface of the conductor from the equator towards the poles and also radially outward. This pulse leaves the simulation domain between $1.5t_c$ and $2t_c$ after which point the inverse square field is established. Fig \ref{chargedynfig} shows the radial electric field throughout this sequence of events.

\subsubsection{Charge Falls in Along Polar Axis}
\label{noisypolar}
In the previous section, we showed that a static field can be established dynamically using {\it PICsar}. However, motion of particles near the polar axis can induce excessive ringing in spherical coordinates. The left panel of Fig \ref{noisyfig} shows the ringing that occurs in spherical coordinates when a particle is shot inward into the central conductor at $v_r=-.95c$ along the polar axis and its partner particle is shot outward along the polar axis at $v_r=.95c$. The plot shows $E_r$ at $t=3.5t_c$ along the polar axis (same elapsed time as panel e of Fig \ref{chargedynfig}). Although the transient has left, there is excessive ringing along the polar axis that does not damp out in time and can inject an unacceptable level of noise into the simulation.

The cause of this ringing in spherical coordinates is due to the increased charge density for a particle near the polar axis in spherical coordinates. This increased charge density is due to the fact that the particle shape function takes the shape of the coordinate grid. In spherical coordinates, the ratio of the volume of a grid cell near the equator to one on the polar axis is $\sim 1/\Delta \theta \approx N_\theta/\pi$. The cells become tiny in volume near the poles and the charge density due to a single particle rises proportionally. Moreover, this problem only becomes {\it worse} with increasing resolution.

We can avoid the concentration of charge for a single particle near the poles altogether by choosing to work in equal area coordinates rather than spherical coordinates. The volume of a grid cell (and hence the charge density of a particle) at constant radius are constant in equal area coordinates (\S \ref{coordsec}), and there is no concentration of charge near the polar axis. 

The right panel of Fig \ref{noisyfig} is the same as the left, but for equal area coordinates rather than spherical coordinates. The ringing near the polar axis resulting from density enhancement in spherical coordinates is eliminated in equal area coordinates. 

\subsection{$E \times B$ drift}
\label{corotationtest}
\begin{figure}[!t]
\centering
\begin{overpic}
	[width=.49\textwidth]{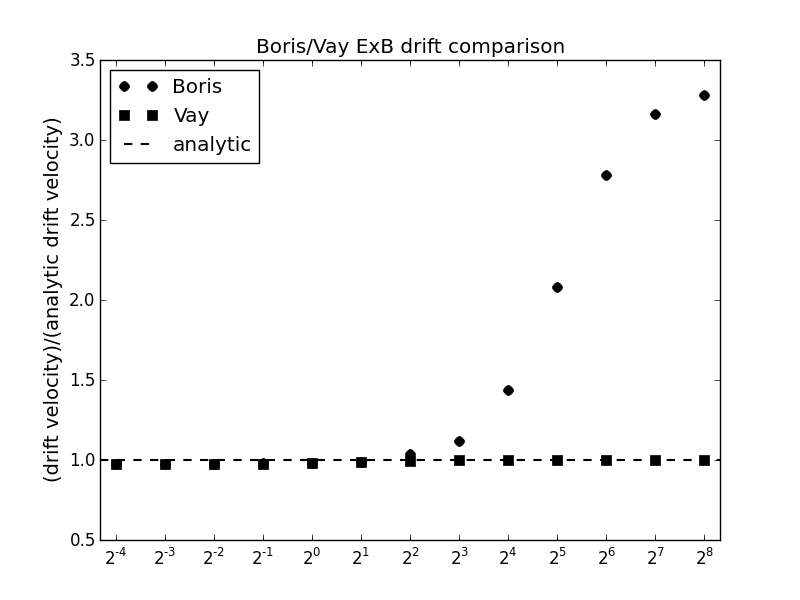}
	\put(50,0){$\omega_c \Delta t$}
\end{overpic}
\caption{$E \times B$ drift motion for the Vay and Boris movers as a function of $\omega_c \Delta t$. The y-axis gives the ratio of the drift velocity measured in the simulation to the analytic drift velocity.}
\label{ExBfig}
\end{figure}

Due to the strength of the electric and magnetic fields in the vicinity of the neutron star, the most important plasma drift to resolve in a pulsar simulation is the $E \times B$ drift \citep{GoldreichJulian}. In a fully kinetic code, it is in general necessary to time-resolve the cyclotron frequency, $\omega_c = qB/mc$, in order to properly capture the $E \times B$ drift motion. However, $B \propto r^{-3}$ for a background dipole magnetic field and can vary by several orders of magnitude over the simulation domain. Thus, if one is most interested in studying the physics at several neutron star radii or further out, then it may be prohibitively expensive to require that the timestep be short enough to resolve the cyclotron frequency at the surface of the neutron star (i.e. $\omega_c(r_0) \Delta t \lesssim 1$).

Fortunately, there exists an algorithm for the particle mover \citep{Vay}, which correctly captures the average $E \times B$ particle drift motion regardless of the value of $\omega_c \Delta t$. This is a special property of the Vay mover, and is not true in general; in particular, it is not true for the commonly used Boris mover \citep{Boris}.

To compare how well the Boris and Vay movers capture the $E \times B$ drift velocity as a function of $\omega_c \Delta t$, we initialize a test particle in background electric and magnetic fields given by
\begin{align}
B_r &= \frac{2\mu}{r^3} \cos \theta ,  &E_r &=  \frac{\mu}{r^2 R_l} \sin^2 \theta \\
B_\theta &= \frac{\mu}{r^3} \sin \theta,  &E_\theta &=-\frac{\mu}{r^2 R_l} \sin 2\theta \nn \\
B_\phi &= 0, &E_\phi &= 0. \nn
\end{align}
The background magnetic field is that of a dipole with magnetic dipole moment $\mu \equiv B_0 r_0^3$, where $r_0$ is the radius of the inner conductor. The background electric field is the ``corotation" electric field, $\bfE = -(\bfOmega \times \bfr) \times \bfB/c$. The light cylinder radius, $R_l$, is the cylindrical radius at which a particle would have to move at the speed of light in order to rotate at the $E \times B$ drift velocity. We use the value $r_0/R_l = .05$ and initialize the test particle at rest in the equatorial plane at a radius of $3 r_0$. 

We then run the simulation using either the Boris or Vay mover and compare the drift velocity of the particle from the simulation with the analytically expected value, $3 \Omega r_0 = .15 c$. Note that although there are gradients of the electric and magnetic fields in the simulation, and there is a centrifugal force on the particle as it drifts azimuthally in a circle around the central conductor, these effects are negligible in our setup compared to the underlying $E \times B$ drift just as in the case of the pulsar.

Fig \ref{ExBfig} shows a plot of the measured $E \times B$ drift velocity normalized by the analytical value of the $E \times B$ drift velocity as a function of $\omega_c \Delta t$. Both the Boris and Vay movers accurately capture the average $E \times B$ drift motion when $\omega_c \Delta t \lesssim 2$. However, for $\omega_c \Delta t \gtrsim 2$ the Boris mover starts to deviate significantly from the analytical solution with increasing $\omega_c \Delta t$, whereas the Vay mover still accurately captures the average $E \times B$ drift motion. Thus, we prefer to use the Vay mover in PIC simulations of pulsars, since one does not need to time-resolve the cyclotron frequency at the surface of the star in order to correctly capture the average $E \times B$ drift motion when using it. 

\section{Aligned Monopole Rotator}
\label{monosec}

\subsection{Analytic Force Free Solution}

We now present the results of aligned monopole rotator simulations using {\it PICsar}. The aligned monopole rotator consists of a central spinning spherical conductor with aligned spin and magnetic axes (we take these to be along the $z$-axis) in a background monopole magnetic field. The force free solution (i.e. when the central spinning conductor is immersed in a conducting plasma with negligible inertia) is useful to consider as a test of the code, because it has an analytic solution given by \citep{Michelmono}:
\begin{align}
\label{monosteady}
B_r &= B_0 \left(\frac{r_0}{r}\right)^2,  &E_r &= 0, &J_r &= -2B_r \Omega \cos \theta \\
B_\theta &= 0,  &E_\theta &=B_\phi, &J_\theta &= 0 \nn \\
B_\phi &= -B_r \frac{R}{R_l} , &E_\phi &= 0, &J_\phi &= 0. \nn
\end{align}
Here $\Omega$ and $r_0$ are the angular frequency and radius of the central conductor, respectively, $B_0$ is the magnetic field at the surface of the conductor, and $R=r\sin\theta$ is the cylindrical radius. The spin of the conductor in the purely radial background magnetic field induces an electric field $\bfE = -(\bfOmega \times \bfr) \times \bfB/c$ and generates a radial current preferentially directed along the $z$-axis. This current creates an azimuthal magnetic field, and the magnetic field lines wind up into an Archimedean spiral in the equatorial plane. The radial and azimuthal magnetic fields are equal in magnitude at the light cylinder.

\subsection{Initialization}

We discuss how to dynamically set up the aligned monopole rotator solution using {\it PICsar}. We start with initial conditions for the fields corresponding to a spinning spherical conductor in vacuum with no plasma in the magnetosphere. We do not model the plasma within the spinning conductor directly as was done by \citet{PhilippovSpitkovsky} but instead split the fields into a pair of ``initial" and ``plasma" fields \citep{Spitkovskythesis}. The initial fields are induced by the rotation of the inner conductor in the background magnetic field and are not due to the particles within the simulation domain; they are analytic, constant in time, and are not used when updating the electromagnetic fields on the grid or applying boundary conditions. On the other hand, the plasma fields are due to the particles within the simulation domain and are updated each timestep. 

The plasma fields are initially zero while the initial fields outside the star ($r > r_0$) are given analytically by 
\begin{align}
\label{initialfields_outside}
B_r &= B_0 \left(\frac{r_0}{r}\right)^2,  &E_r &=  -2 B_0 \left(\frac{r_0^4}{r^3R_l}\right) \cos \theta  \\
B_\theta &= 0,  &E_\theta &=-B_0\left(\frac{r_0^4}{r^3R_l}\right) \sin \theta \nn \\
B_\phi &= 0, &E_\phi &= 0. \nn
\end{align}
Inside the star ($r < r_0$) the initial fields are given by
\begin{align}
\label{initialfields_inside}
B_r &= B_0 \left(\frac{r_0}{r}\right)^2,  &E_r &=  0  \\
B_\theta &= 0,  &E_\theta &=-B_0\left(\frac{r_0^2}{rR_l}\right) \sin \theta \nn \\
B_\phi &= 0, &E_\phi &= 0, \nn
\end{align}
and the electric field is simply the corotation electric field for a monopole magnetic field geometry. The reason we need to specify the fields inside the star is because we can only delete particles that penetrate far enough into the inner conductor that their shape function lies entirely within the conductor (see description of ``hanging charge" in \S \ref{pbc}). Thus, particles inside the inner conductor which do not penetrate far enough to be deleted move under the influence of corotation fields. 

By splitting the fields into vacuum and plasma fields, we are able to retain the simple boundary condition for an unmagnetized non-rotating conductor ($\bfE = \bfB = 0$) at the inner radius of the simulation domain for the plasma fields. At the outer radius, we use the radiation absorbing boundary conditions described in \S \ref{fbc}.

By the linearity of Maxwell's equations, the total field is simply the sum of the vacuum and plasma fields. This total field is used, for example, when computing the Lorentz force felt by the particles and injecting pairs into the simulation domain, which we discuss in the next section.

To avoid a large initial transient, we spin up the star on a timescale $t_\text{spin} \sim \Omega^{-1}$ \citep{Spitkovskythesis}. During the spinup phase, $\Omega$, the angular velocity of the central conductor increases linearly to its final constant value.

\begin{figure*}[!t]
\centering
\subfigure{\begin{overpic}
		  [width=.3025\textwidth]{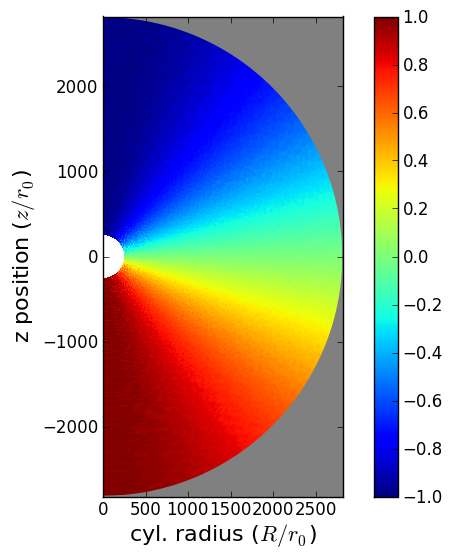}
		  \put(3,92){\normalsize a)}
		  \put(42,90){\normalsize$t=2P$}
		  \put(48,16){\large$r^2J_r$}
		  \end{overpic}}
\subfigure{\begin{overpic}
		  [width=.4875\textwidth]{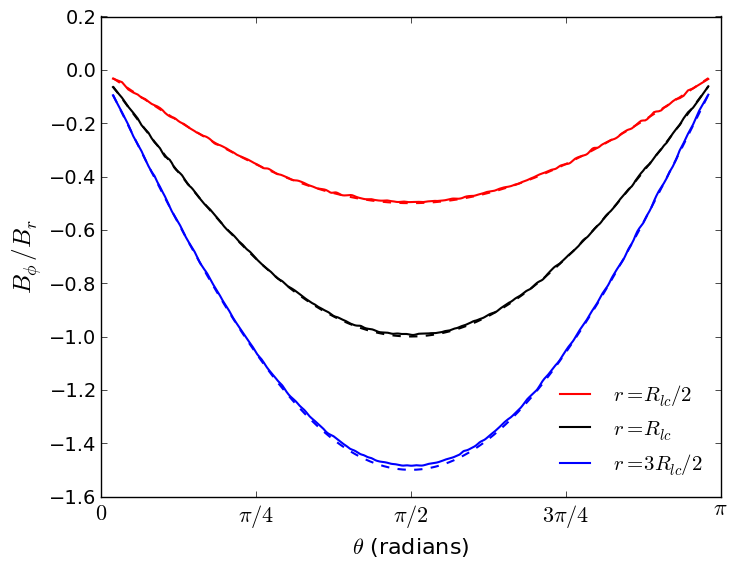}
		  \put(1,70){\normalsize b)}
		  \put(83,68.5){\normalsize$t=2P$}
		  \end{overpic}}
\subfigure{\begin{overpic}
		  [width=.2425\textwidth]{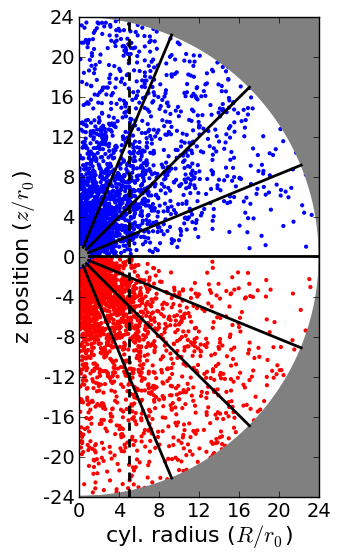}
		  \put(1,91){\normalsize c)}
		  \put(39,90){\normalsize$t=2P$}
		  \end{overpic}}
\subfigure{\begin{overpic}
		  [width=.5175\textwidth]{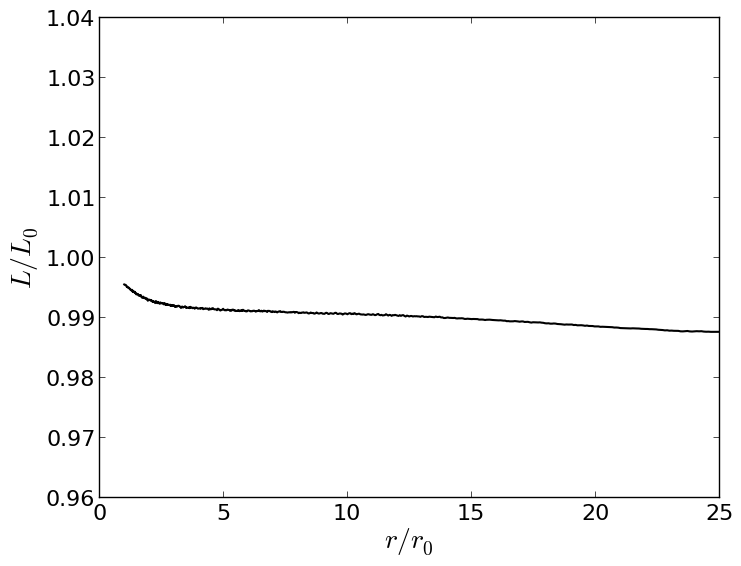}
		  \put(0,70){\normalsize d)}
		  \put(80.5,68){\normalsize$t=2P$}
		  \end{overpic}}
\caption{Aligned monopole rotator simulation after two rotation periods. a) $r^2J_r$ normalized to the analytical value of $r^2J_r$ at the pole. b) $B_\phi/B_r$ as a function of meridional angle at three different radii. Solid lines are simulation results and dashed lines are the analytic solution. c) Downsampled charge density distribution as represented by particles( blue/red -- positive/negative particles). Solid black lines are magnetic field lines in the meridional plane. d) Simulation spindown luminosity normalized by the analytic spindown luminosity as a function of radius.}
\label{monofig}
\end{figure*}
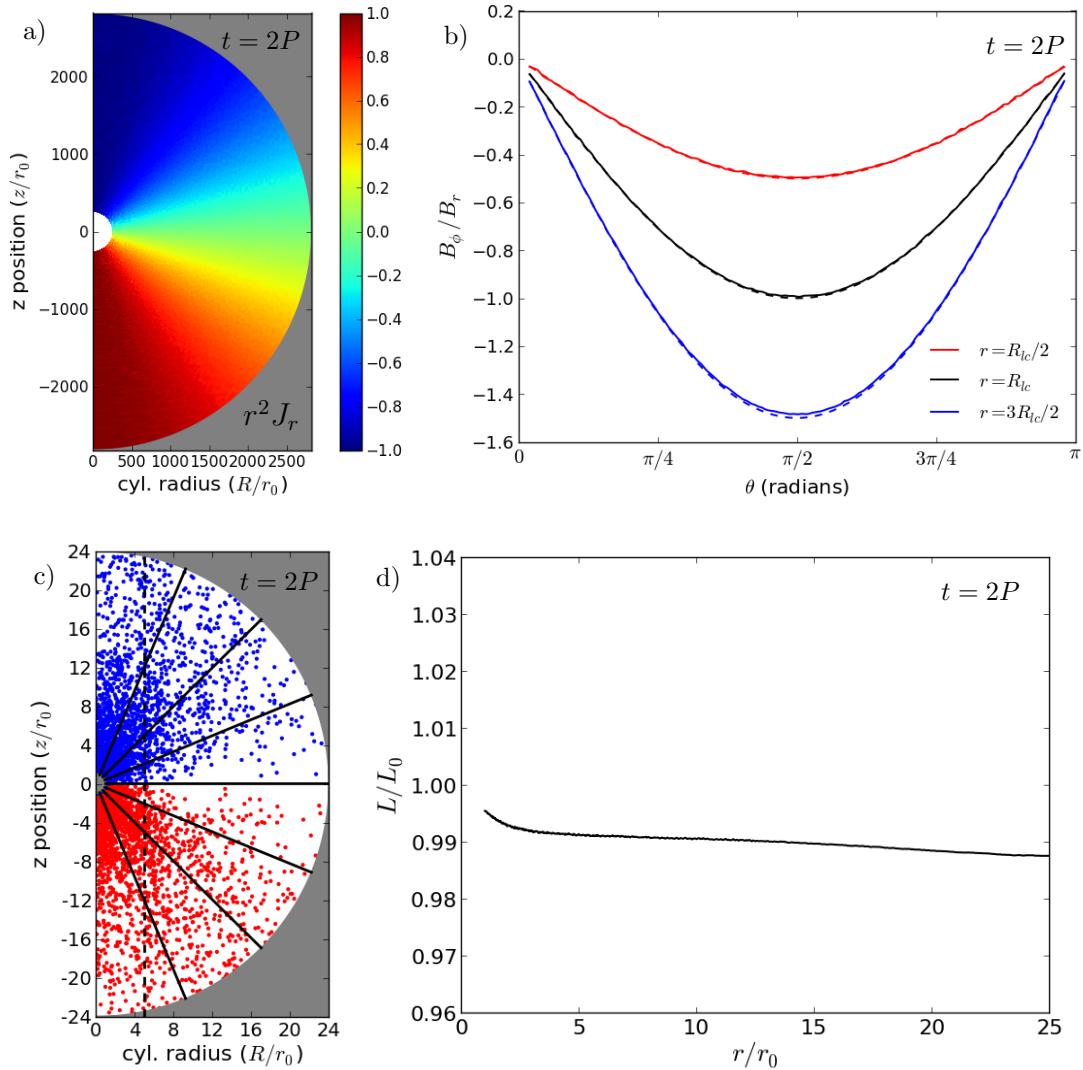

\subsection{Surface Charge Injection}
\label{injsec}

Initially our simulation contains no particles so the plasma field is zero. However, because the star (central conductor) is rotating in a magnetic field, a charge is induced on its surface. As a result, there is a jump in the analytic fields at the surface of the star from corotation fields inside to vacuum fields outside. 

Our prescription for filling the magnetosphere with plasma is to locally ``release" the accumulated surface charge in the form of particles into the simulation domain. The accumulated charge is calculated based on the jump in the radial component of the electric field across the conductor; it is injected at rest each timestep just above the surface of the star. However, injecting all of the surface charge each timestep can potentially result in virtual cathode oscillations. This is because the particles have a finite width shape function that is broadened by filtering. Therefore, it effectively takes more than one timestep for a newly injected particle to entirely carry away the charge from the surface. 

In order to avoid virtual cathode oscillations and generate a steady flow from the surface of the star, one can inject a fraction $f_\text{inj}$ of the accumulated surface charge on the conductor each timestep, where $0 < f_\text{inj} \le 1$. Our general ansatz for charge injection at the surface of the star is then given by 
\ba
\label{volinjeq}
\dot{n_j} = \frac{f_\text{inj} \sigma_j}{q} dA_j.
\ea
Here $\dot{n_j}$ is the number of pairs injected into the $j$-th cell (meridional direction) which is adjacent to the inner conductor, $q$ is the particle charge, and $dA_j$ is surface area of the cell on the face of the conductor. The pairs are created with opposite charges at the same position, which means we do not have to solve Poisson's equation. Also, the fact that there is a surface charge implies a nonzero value of $\bfE \cdot \bfB$ above the star ($\bfE \cdot \bfB = 0$ inside), meaning one sign of charge will be pulled into the star and the other will accelerate away from the star and into the magnetosphere.

The simulation results obtained using our injection algorithm depend weakly on the value of $f_\text{inj}$. Smaller values of $f_\text{inj}$ are more stable (in the sense that they don't exhibit virtual cathode oscillations), but values of $f_\text{inj}$ closer to unity are more accurate when compared with the analytic solution. Smoothing the value of $\sigma_j$ along the surface of the star in the meridional direction can help increase stability, but the smoothing length should be less than the smallest meridional scale of interest.

The reason the simulation results depend weakly on $f_\text{inj}$ is that the prescription (\ref{volinjeq}) exponentially decreases the charge on the surface of the star on a characteristic damping timescale, which is $1/f_\text{inj}$ timesteps. As long as this timescale is shorter than the relevant physical timescales of interest, the result will approximate the physical solution with no surface charge on the star (zero work function for particles to escape). 

We emphasize that our injection algorithm is a {\it surface} injection algorithm, and that for steady injection it leads to a space charge separated plasma. For simulations of the aligned dipole rotator, the surface injection algorithm needs to be coupled with a pair production prescription in order to generate the force free solution. The particles emitted from the surface would then form the seeds for the pair production cascade. However, surface injection with no pair production is sufficient for generating the force free solution of the aligned monopole rotator, because all the monopolar field lines are open.

\subsection{Simulation Results}

We now discuss the simulation results for the aligned monopole rotator. For a space charge separated plasma, the force free solution implies that the velocity of the particles is equal to the speed of light. Thus, we will obtain the force free solution in the limit $\gamma \rightarrow \infty$. Consequently, our results will depend weakly on the strength of the electromagnetic fields as long as particles are accelerated to highly relativistic velocities. Additionally, the magnetic field should be strong enough that the Larmor radius of the particles is small compared with the characteristic physical length scale of the problem which is the simply the spherical radius in this case. The median value of the particle gamma factor is $\gamma \approx 28$ in our simulation, and the value of the Larmor radius is less than a grid cell at the surface of the star; the latter necessitates the use of the Vay mover (\S \ref{corotationtest}).

For the coordinate system in the simulation, we use logarithmic and equal area coordinates in the radial and meridional directions respectively. A set of relevant numerical parameters is given in Table \ref{paramtable}. We also mention that both the positively and negatively charged particles have the same mass and charge so we are simulating a pair plasma. 

\begin{table}
\caption{Numerical parameters for monopole simulation.} 
\label{paramtable}
\centering
\begin{tabular}{|l|l|} 
\hline
grid dimensions & $N_r \times N_\theta =  1024 \times 512$ \\
number of particles & $\approx 2 \times 10^7$ \\
inner conductor radius & $r_0 = 256$ \\
outer conductor radius & $r_\text{max} \approx 54 r_0$ \\
light cylinder radius & $R_l = 5 r_0$ \\
Courant number (speed of light) & $c = .6$ \\
simulation duration & 2 rotational periods \\
current filtering & $N_\text{filt} = 3$ \\
injection multiplier & $f_\text{inj}$ = .5 \\
spinup time & $t_\text{spin} = .5$ rotational periods \\
\hline
\end{tabular}
\end{table}

Fig \ref{monofig} shows the results of the aligned monopole rotator simulation with surface injection after 2 rotation periods of the central conductor when a dynamic steady state has been set up. This steady state is very nearly the force-free solution with a thin accelerating region of thickness a few cells near the surface of the star, where injected particles accelerate to relativistic velocities.

Fig \ref{monofig}a shows $r^2J_r$ normalized by $r_0^2 J_0 = 2 B_0 \Omega r_0^2$, where $J_r$ is the current density in the meridional plane. The analytical expression for $r^2J_r$ in the case of the monopole is $r^2 J_r = -J_0 \cos \theta$ (equation \ref{monosteady}).

Fig \ref{monofig}b shows a plot of $B_\phi/B_r$ as a function of $\theta$ at three different radii $r = R_{l}/2$ (red), $r = R_{l}$ (black), and $r=3R_{l}/2$ (blue). The dashed lines are the analytical solutions according to equations (\ref{monosteady}), and the solid lines are simulation results. Note that $B_\phi = 0$ initially and that the azimuthal magnetic field is generated exclusively by the current due to the particles.

Fig \ref{monofig}c shows a heavily downsampled image of the particles in the simulation. The surface injection algorithm has produced a space charge separated plasma with the positively charged particles (red) and negatively charged particles (blue) occupying distinct regions of the simulation domain. Although each particle represents a charge annulus, the downsampling has been done in such a way that the clustering of particles corresponds to the particle density in a 3D simulation.

Fig \ref{monofig}d shows the simulation spindown luminosity (Poynting flux integrated over the surface of a sphere) normalized by the analytical value of the spindown luminosity as a function of radius out to $25 r_0$. In Gaussian units, the analytical value of the spindown luminosity for the force-free aligned monopole rotator is
\ba
L_0 = \frac{2c}{3}\left(\frac{B_0 r_0^2}{R_{l}}\right)^2.
\ea
The value calculated from the simulation is within $\sim 1\%$ of the analytical value and is approximately constant with radius as expected. 

\section{Discussion}
\label{discussion_sec}

\subsection{{\it PICsar} speedup}
\label{PICsar_speedup}

We have shown that it is possible to accurately simulate the magnetosphere of the aligned rotator using a 2.5D axisymmetric PIC code. One reason for using a 2.5D axisymmetric code versus a Cartesian code is that a 2.5D axisymmetric code is orders of magnitude faster than a 3D Cartesian code.
 
We can roughly estimate the speedup of a 2.5D axisymmetric code versus a 3D Cartesian code by comparing the number of cells in the two codes for the same simulation volume and assuming the same number of particles per cell for both simulations. To within factors of order unity that arise from different treatments of the boundary conditions, coordinate transformations etc., the speedup is simply the ratio of the number of cells in the axisymmetric code to the number of cells in the Cartesian code. Assuming the $(\xi,\theta_A)$ coordinate system for the axisymmetric code and cubical cells for the Cartesian code, the speedup ratio is
\ba
\label{speedup1}
\frac{\text{CPU time 3D Cartesian}}{\text{CPU time 2.5D axisym}} \sim \frac{(r_\text{max}/r_0)^{2} r_\text{max}}{1+\ln(r_\text{max}/r_0)},
\ea
where $r_0$ and $r_\text{max}$ are then inner/outer conductor radii both measured in cells. For $r_\text{max}/r_0 \gg 1$ and $r_\text{max} \gg 1$, the 2.5D axisymmetric code provides an {\it orders of magnitude} increase in speed over a 3D Cartesian code.

\subsection{{\it PICsar} 3D}

Up to now we have only discussed methods for simulating the aligned rotator using a 2.5D axisymmetric code. However, the algorithms we have presented transfer naturally to the 3D case, as long as the 3D computational grid can be deformed to a 3D Cartesian grid without breaking the grid lines. One example of such a coordinate system, which is well-suited to pulsar simulation is the cubed sphere coordinate system. 

The cubed sphere coordinate system is essentially a Cartesian coordinate system that has been deformed to fit the surface of a sphere. Although there would be extra edge and corner cases to consider, a fully 3D simulation in cubed sphere coordinates would still be orders of magnitude faster than a 3D Cartesian simulation. Using the same estimation method as in \S \ref{PICsar_speedup}, one can estimate the speedup of a 3D cubed sphere code (logarithmic radial coordinate) versus a 3D Cartesian code as
\ba
\label{speedup2}
\frac{\text{CPU time 3D Cartesian}}{\text{CPU time 3D cube sph}} \sim \frac{\left(r_\text{max}/r_0\right)^3}{1+\ln\left(r_\text{max}/r_0\right)}.
\ea
Again, the cubed sphere code is much faster than the Cartesian code in the limit $r_\text{max}/r_0 \gg 1$.

The reason for the large speedup factor in 3D is due to the fact that stretchable coordinates body-fitted to the sphere are more naturally suited to the pulsar problem than a Cartesian grid. Whereas 3D Cartesian coordinates have a constant resolution over the whole simulation grid, logarithmic radial coordinates have a resolution which effectively varies as $\propto 1/r$ (in each dimension) so the resolution is concentrated close to the surface of the star. This is typically what one desires, especially when putting the outer boundary far enough away to eliminate any adverse effect from it on the physics of interest.

\subsection{Future Science}
Because {\it PICsar} is a particle-mesh code, it allows investigation of phenomena that are difficult to probe via purely mesh based approaches such as force-free or MHD. Examples of such phenomena include pair production, particle acceleration, and kinetic processes. 

One area that remains relatively unexplored in global simulations is pair production and particle acceleration in magnetospheric vacuum gaps. Although 1D PIC simulations of pair producing vacuum gaps have been carried out \citep{Timokhin_paircascades,TimokhinArons}, these simulations are divorced from the global structure of the magnetosphere. 2.5D axisymmetric PIC simulations allow deeper insight into how vacuum gaps operate in the global magnetospheric context \citep{chenbeloborodov}; {\it PICsar} is well-suited to such investigations. 

A second area to which {\it PICsar} can be applied is the study of current sheet instabilities. \citet{ZenitaniHoshino2007,ZenitaniHoshino2008} have shown using linear theory and local PIC simulations that an electron positron current sheet without a strong guide field (the relevant case for a pulsar beyond the light cylinder) should be susceptible to the relativistic drift kink instability (RDKI). Moreover, since the RDKI has unstable modes that lie in the $r-\theta$ plane, observing RDKI in a 2.5D axisymmetric simulation is realistic, and it has already been noted by \citet{PhilippovSpitkovsky} in their 3D PIC simulations of the aligned rotator.

A third area that can be investigated with {\it PICsar} is particle acceleration due to reconnection. Reconnection in the current sheet occurring near the pulsar termination shock has been suggested as the source of particle acceleration that generates gamma ray flares in the Crab nebula \citep{Uzdenskyetal,CeruttiUzdenskyBegelman}. It would be interesting to investigate whether such flaring events occur spontaneously in a global pulsar simulation and lead to particle acceleration.

\subsection{Summary}
We have presented a new 2.5D axisymmetric relativistic, electromagnetic, charge conservative PIC code that is well-suited to simulations of the aligned dipole rotator. The new code is parallelized using MPI and implements stretchable coordinates in both the meridional and radial directions that are body-fitted to the surface of a sphere. Stretch factors for the coordinates are used that minimize particle shot noise by accommodating a greater number of particles per cell for a fixed total number of particles. Also, because the coordinate system follows the surface of a sphere, the conducting boundary condition on the surface of the star is trivial to implement, and we do not need to simulate the plasma in the neutron star interior. 

In order to set up a steady state dynamically, the code implements a radiation absorbing outer layer inside an outer conducting shell. The radiation absorbing layer damps electromagnetic waves propagating through it, and the spherical conducting shell facilitates a surface current, redistributing charge and preventing local charge accumulation at the outer boundary.  

For the particle push step, we prefer to use the Vay algorithm over the more commonly used Boris algorithm. The Vay algorithm exactly captures the dominant $E \times B$ drift regardless of whether the cyclotron frequency is resolved in time.  

Finally, we have presented an algorithm for charge injection that releases the accumulated charge on the surface of the star into the simulation domain. The new algorithm injects pairs near the surface in proportion to the local value of the surface charge on the star. For the aligned monopole rotator, it naturally generates a solution which is everywhere force free, except for thin acceleration gaps, where particles are accelerated to relativistic velocities. These acceleration gaps are directly analogous to the vacuum gaps occurring in pulsar magnetospheres that are the sites of pair creation.

The author would like to thank Anatoly Spitkovsky for suggesting the project, for laying the foundation that enabled this research to be possible, and for many useful discussions. Additionally, the author would like to thank Lorenzo Sironi and Jon Arons for useful discussions. 

\bibliography{pulsar}
\appendix

\section{Field Solve in Body-Fitted Coordinates}
\label{field_solve_app}

We show explicitly how to solve Maxwell's equations in conservative form in body-fitted coordinates. We begin by rewriting equations (\ref{maxtd}) in integral form
\ba
\label{intmaxwell}
\frac{\partial}{\partial t} \iint \bfB \cdot \bfdA &=& - c\oint \bfE \cdot \bfdl \\
\frac{\partial}{\partial t} \iint \bfE \cdot \bfdA &=&  c\oint \bfB \cdot \bfdl - \oiint \bfJ \cdot \bfdA. \nn
\ea

Using the definition (\ref{rtdefs}), equations (\ref{intmaxwell}) become to second order in space
\ba
\label{discreteeq}
\frac{\partial B_r^{ij}}{\partial t} dA_{\theta \phi}(r_i,[\theta_j,\theta_{j+1}]) &=& c\left[E_\phi^{ij}dL_\phi(r_i,\theta_j) - E_\phi^{i j+1}dL_\phi(r_i,\theta_{j+1})\right] \\
\frac{\partial B_\theta^{ij}}{\partial t} dA_{r \phi}([r_i,r_{i+1}],\theta_j) &=& c\left[E_\phi^{i+1j}dL_\phi(r_{i+1},\theta_j) - E_\phi^{i j}dL_\phi(r_i,\theta_j)\right] \nn \\
\frac{\partial B_\phi^{ij}}{\partial t} dA_{r \theta}([r_i,r_{i+1}],[\theta_j,\theta_{j+1}]) &=& c\Big\{ \Big[E_\theta^{ij}dL_\theta(r_{i},[\theta_j,\theta_{j+1}]) - E_\theta^{i+1 j}dL_\theta(r_{i+1},[\theta_j,\theta_{j+1}])\Big]  \nn \\
				&-& \Big[E_r^{ij} dL_r([r_i,r_{i+1}]) - E_r^{i j+1}dL_r([r_i,r_{i+1}])\Big] \Big\} \nn \\
\left[\frac{\partial E_r^{ij}}{\partial t} + J_r^{ij}\right] dA_{\theta \phi}(\ovr_i,[\ovt_{j-1},\ovt_j]) &=& c\left[B_\phi^{ij}dL_\phi(\ovr_i,\ovt_j) - B_\phi^{i j-1}dL_\phi(\ovr_i,\ovt_{j-1})\right] \nn \\
\left[\frac{\partial E_\theta^{ij}}{\partial t} + J_\theta^{ij}\right] dA_{r \phi}([\ovr_{i-1},\ovr_i],\ovt_j) &=& c\left[B_\phi^{i-1j}dL_\phi(\ovr_{i-1},\ovt_j) - B_\phi^{i j}dL_\phi(\ovr_i,\ovt_j)\right] \nn \\
\left[\frac{\partial E_\phi^{ij}}{\partial t} + J_\phi^{ij}\right] dA_{r \theta}([\ovr_{i-1},\ovr_i],[\ovt_{j-1},\ovt_j]) &=& c\Big\{ \Big[B_\theta^{ij}dL_\theta(\ovr_{i},[\ovt_{j-1},\ovt_j]) - B_\theta^{i-1 j}dL_\theta(\ovr_{i-1},[\ovt_{j-1},\ovt_j])\Big]  \nn \\
				&-& \Big[B_r^{ij} dL_r([\ovr_{i-1},\ovr_i]) - B_r^{i j-1}dL_r([\ovr_{i-1},\ovr_i])\Big] \Big\} \nn. 
\ea

To reduce the number of parentheses and symbols, we have defined
\ba
\label{rtdefs}
r_i &\equiv& f_r(i), \ \ \ \ovr_i \equiv f_r(i+1/2) \\
\theta_j &\equiv& f_\theta(j), \ \ \ \ovt_j \equiv f_\theta(j+1/2), \nn
\ea
where $f_r$ and $f_\theta$ are the coordinate stretching functions introduced in \S \ref{coordsec}. Also, we have introduced the area elements, $dA_{\theta \phi}, \ dA_{r \phi}, \ dA_{r\theta}$, and the line elements $dL_r, \ dL_\theta, \ dL_\phi$, defined as
\begin{align}
\label{linearea}
dA_{\theta \phi}(r,[\theta_a,\theta_b]) &\equiv \int_{0}^{2\pi} \int_{\theta_a}^{\theta_b} r^2 \sin \theta  d\theta d\phi, &dL_r([r_a,r_b]) &\equiv \int_{r_a}^{r_b} dr \\
&= 2\pi r^2(\cos\theta_a - \cos\theta_b), & &= r_b-r_a\nn \\
dA_{r \phi}([r_a,r_b],\theta) &\equiv  \int_{0}^{2\pi} \int_{r_a}^{r_b} r \sin \theta  dr d\phi, &dL_\theta(r,[\theta_a,\theta_b]) &\equiv \int_{\theta_a}^{\theta_b} r d\theta \nn \\ 
&= \pi(r_b^2-r_a^2)\sin \theta, & &= r(\theta_b - \theta_a) \nn \\
dA_{r \theta}([r_a,r_b],[\theta_a,\theta_b]) &\equiv \int_{\theta_a}^{\theta_b} \int_{r_a}^{r_b} r  dr d\theta, &dL_\phi(r,\theta) &\equiv \int_{0}^{2 \pi} r \sin \theta d\phi \nn \\ 
&= \frac{1}{2}(r_b^2-r_a^2)(\theta_b-\theta_a), & &= 2 \pi r \sin \theta \nn.
\end{align}
Equations (\ref{discreteeq}) together with the area and line elements (\ref{linearea}) are generalizations of the update equations given by \citet{THREDS} for spherical coordinates and allow for coordinate stretching in $\theta$ and $r$. The time discretization of equations (\ref{discreteeq}) in {\it PICsar} is the standard second order leapfrog for the Yee method.

Due to the fact that the meridional boundaries lie on the polar axis, the meridional boundary conditions for $B_\theta$ and $E_\phi$ are simply $B_\theta=E_\phi=0$, which is true in axisymmetry. $E_r$ has to be updated in a special way on the polar axis, and the update equations for $E_r$ on the polar axis ($j=1$ or $j=N_\theta$) are
\ba
\label{erpolarupdate}
\left[\frac{\partial E_r^{i,1}}{\partial t} + J_r^{i,1}\right] dA_{\theta \phi}(\ovr_i,[0,\ovt_1]) &=& cB_\phi^{i,1}dL_\phi(\ovr_i,\ovt_1) \\
\left[\frac{\partial E_r^{i,N_\theta}}{\partial t} + J_r^{i,N_\theta}\right] dA_{\theta \phi}(\ovr_i,[\ovt_{N_\theta-1},\pi]) &=& -cB_\phi^{i,N_\theta-1}dL_\phi(\ovr_i,\ovt_{N_\theta-1}) \nn \\
\ea
This can be derived in the same way as the update equations for the main body of the simulation domain (equations (\ref{discreteeq})), by taking line and area integrals around the polar cap \citep{THREDS}. Because $E_\theta$, $B_\phi$, and $B_r$ are all located at half integer meridional coordinates, they do not lie on the meridional boundaries, and thus no boundary condition in the meridional direction needs to be applied for these field components.

\section{Current Filtering}
\label{filterapp}

We now provide a more detailed description of current filtering. For the 1-2-1 digital filter used in {\it PICsar}, the current filtering step can be written as
\ba
\label{filtereq}
\overline{J}^{ij}_* &=&  \sum_{l = -1}^1 \sum_{k = -1}^1 W_k W_l J^{i+k,j+l}_* \\
W_m &=& \left \{\begin{array}{ll}
1/4, & m = \pm 1 \\
1/2, & m = 0
\end{array}
\right. 
\ea
where $J$ is the unfiltered current, $\overline{J}$ is the filtered current, $W_m$ are the weights for the 1-2-1 digital filter, and the star notation just means that the formula holds for all vector components. However, care must be taken when applying the filter close to the boundaries of the simulation domain.

\subsection{Filtering Near the Polar Axis}

The current filtering prescription (\ref{filtereq}) must be modified near the polar axis to prevent current from being filtered outside the simulation domain through the meridional boundaries. This modification results in a``remapping" of the current which would be filtered outside the simulation domain back into the simulation domain. Current remapping near the polar axis modifies the filtering formula (equation (\ref{filtereq})) and weights in the following way:
\ba
\overline{J}_*^{ij} &=&  \sum_{l = -1}^1 \sum_{k = -1}^1 W_k V_l J_*^{i+k,j+l} \\
* = r,\phi &\longrightarrow&
\begin{array}{l}
\left \{
	\begin{array}{ll}
	    j = 1, &V_l = 
		\left \{
			\begin{array}{ll}
					0, & l = -1 \\
					1/2, & l = 0 \\
					1/4, & l =1
			\end{array}
		\right. \\
	j = 2, &V_l =
	\left \{
	    \begin{array}{ll}
		    1/2, & l = -1 \\
		    1/2, & l = 0 \\
		    1/4, & l =1
	    \end{array}
	\right. \\
	j = N_\theta-1, &V_l = 
	\left \{
		\begin{array}{ll}
		    1/4, & l = -1 \\
		    1/2, & l = 0 \\
		    1/2, & l =1
		\end{array}
	\right. \\
        j = N_\theta, &V_l =
	\left \{
	    \begin{array}{ll}
		    1/4, & l = -1 \\
		    1/2, & l = 0 \\
		    0, & l =1
	    \end{array}
	\right.
	\end{array} 
\right.
\end{array} \nn \\
* = \theta &\longrightarrow&
\begin{array}{l}
\left \{
	\begin{array}{ll}
	 j = 1, &V_l = 
	\left \{
	   \begin{array}{ll}
		    0, & l = -1 \\
		    1/4, & l = 0 \\
		    1/4, & l =1
	   \end{array}
	\right. \\
	j = N_\theta-1, &V_l =
	\left \{
	    \begin{array}{ll}
		    1/4, & l = -1 \\
		    1/4, & l = 0 \\
		    0, & l =1
	    \end{array}
	\right. 
	\end{array} 
\right.
\end{array} \nn 
\ea
Here, $W_k$ is the same ``1-2-1" weighting factor as given in equation (\ref{filtereq}), and the ``$* = $" notation indicates to which component of $J$ we are referring. The reason why the current remapping differs for the $* = r,\phi$ and the $* = \theta$ components is twofold. First, $J_\theta$ is defined at half-integer $j$-coordinates, whereas $J_r, J_\phi$ are defined at integer $j$-coordinates (equation (\ref{offseteq})). Second, the remapped current comes in with a positive sign when it is folded back into the simulation domain for the $J_r, J_\phi$ components, whereas it comes in with a negative sign for the $J_\theta$ component. This produces the effect of a reflection of the particle shape function off the polar axis.

\subsection{Charge Conservation with Current Filtering}

We show that digitally filtering the current after it is deposited to the grid using a charge conservative deposition scheme does not compromise charge conservation. Rather, it results in a modified charge conservative deposition scheme for which the particle shape function is the same as in the original deposition scheme (no filtering) but which has been filtered $N_\text{filt}$ times.

For simplicity, we will consider the 1D case, which simplifies the algebra. In 1D, we take the (only) grid coordinate $i$ to be in the $x$-direction. In this case equation \ref{filtereq} for the filtered current becomes
\ba
\label{onepassfilter}
\overline{J}^{i} &=&  \sum_{k = -1}^1 W_k J^{i+k},
\ea
where $J$ is understood to be current in the $x$-direction, so we suppress the coordinate subscript. Equation (\ref{onepassfilter}) is for one pass of the digital filter. The effect of applying the filter to the current multiple times yields the following expression for the filtered current
\ba
\overline{J}^{i} &=& \sum_{n = -1}^1 W_n ... \sum_{l = -1}^1 W_l \sum_{k = -1}^1 W_k J^{i+k+l+...+n} \nn \\
&=& \sum_{n = -1}^1 ... \sum_{l = -1}^1 \sum_{k = -1}^1 W_n ... W_l W_k J^{i+k+l+...+n}. 
\ea

Because the current deposition scheme is charge conservative by assumption, the change in the charge enclosed by the $i$-th cell from one timestep to the next is given by 
\ba
\Delta Q^i \equiv J^{i-1} - J^i,
\ea
where $J$ is the unfiltered current. Using linearity, we can then write
\ba
\overline{J}^{i-1} - \overline{J}^{i}  &=& \sum_{n = -1}^1 ... \sum_{l = -1}^1 \sum_{k = -1}^1 W_n ... W_l W_k \Delta Q^{i+k+l+...+n}. 
\ea
In order for the filtering to be charge conservative, {\it we demand} that 
\ba
\overline{\Delta Q}^i = \overline{J}^{i-1} - \overline{J}^i,
\ea
which implies that 
\ba
\label{multfiltchargedelta}
\overline{\Delta Q}^i  &=& \sum_{n = -1}^1 ... \sum_{l = -1}^1 \sum_{k = -1}^1 W_n ... W_l W_k \Delta Q^{i+k+l+...+n}. 
\ea

Assuming we have used proper initial conditions that are consistent with charge conservation, equation (\ref{multfiltchargedelta}) implies that
\ba
\label{multfiltcharge}
\overline{Q}^i  &=& \sum_{n = -1}^1 ... \sum_{l = -1}^1 \sum_{k = -1}^1 W_n ... W_l W_k Q^{i+k+l+...+n},
\ea
where $Q$ and $\overline{Q}$ are now the unfiltered and filtered charge densities at the current time step, respectively. Thus, the charge density in a given cell is simply the filtered charge density implied by the charge conservative deposition scheme. Since the charge density in a PIC code is due to a summation over individual particles, equation (\ref{multfiltcharge}) will be satisfied in general (i.e. current filtering will be charge conservative), if the shape function of the particles implied by the current deposition scheme has been filtered the same number of times as the current.

\end{document}